\documentclass[9pt]{extarticle}
\usepackage{spconf,amsmath,graphicx}
\usepackage{amssymb}
\usepackage{multirow}
\usepackage{booktabs}
\usepackage{ctable}
\usepackage{stix}
\usepackage{cite}

\title{MossFormer2: Combining Transformer and RNN-Free Recurrent Network for Enhanced Time-Domain Monaural Speech Separation}
\name{%
\begin{tabular}{@{}c@{}}
Shengkui Zhao  \qquad 
Yukun Ma\qquad 
Chongjia Ni \qquad 
Chong Zhang \\  
Hao Wang \qquad  
Trung Hieu Nguyen \qquad  
Kun Zhou \qquad
Jia Qi Yip \qquad
Dianwen Ng \qquad
Bin Ma
\end{tabular}}
\address{Alibaba Group\\
	\{shengkui.zhao, b.ma\}@alibaba-inc.com\\}
\begin{document}
%
\maketitle
\begin{abstract}
Our previously proposed MossFormer has achieved promising performance in monaural speech separation. However, it predominantly adopts a self-attention-based MossFormer module, which tends to emphasize longer-range, coarser-scale dependencies, with a deficiency in effectively modelling finer-scale recurrent patterns. In this paper, we introduce a novel hybrid model that provides the capabilities to model both long-range, coarse-scale dependencies and fine-scale recurrent patterns by integrating a recurrent module into the MossFormer framework. Instead of applying the recurrent neural networks (RNNs) that use traditional recurrent connections, we present a recurrent module based on a feedforward sequential memory network (FSMN), which is considered "RNN-free" recurrent network due to the ability to capture recurrent patterns without using recurrent connections. Our recurrent module mainly comprises an enhanced dilated FSMN block by using gated convolutional units (GCU) and dense connections. In addition, a bottleneck layer and an output layer are also added for controlling information flow.
The recurrent module relies on linear projections and convolutions for seamless, parallel processing of the entire sequence. The integrated MossFormer2 hybrid model demonstrates remarkable enhancements over MossFormer and surpasses other state-of-the-art methods in WSJ0-2/3mix, Libri2Mix, and WHAM!/WHAMR! benchmarks.

\end{abstract}   
\begin{keywords}
speech separation, transformer, attention, convolution, recurrent, deep learning
\end{keywords}
\section{Introduction}
\label{sec:intro}

Monaural speech separation, referring to  the estimation of individual speech sources from a single-channel mixture of speech sources, is a fundamental and crucial endeavor for numerous downstream speech applications \cite{Lam2019W, Neumann2020K}. 
Lately, substantial enhancements in performance have been achieved through the utilization of end-to-end deep learning models. In general, speech separation techniques can be broadly categorized into two main approaches based on the input types: the time-frequency (T-F) domain approach and the time domain approach. Until recently, significant advancements have been achieved in the T-F domain approach \cite{Yang2022L, wang2023tfgridnet}. The progress can be mainly attributed to the complex-valued T-F representation and the scanning of multiple cross-frame and cross-frequency paths. However, the present model architectures heavily depend on the bidirectional LSTM network, resulting in substantial computational demands and slow inference speed. As a consequence, their scalability is constrained, particularly for more demanding scenarios.

Meanwhile, the time-domain techniques \cite{Luo2019N,Luo2020Z,Nachmani2020Y,Chen2020Q,Zeghidour2021D,Subakan2021M} have become the dominant approach following the introduction of TasNet \cite{Luo2018M}. Generally, the time-domain methodology employs an encoder-separator-decoder framework. It has been observed that utilizing a smaller kernel size in the encoder often results in enhanced separation performance \cite{Luo2019N}. However, this practice also leads to significantly longer sequences, which poses efficiency challenges during sequential processing. The dual-path method proposed by Luo et al.\cite{Luo2020Z} tried to address this issue by partitioning the sequence into smaller chunks and then going through alternative intra-processing and inter-processing steps. Subsequently, numerous studies have embraced the dual-path approach, focusing on the development of separator architectures. These include recurrent models such as DPRNN  \cite{Luo2020Z} and Gated DPRNN \cite{Nachmani2020Y}, as well as Transformer-based models like DPTNet \cite{Chen2020Q} and SepFormer \cite{Subakan2021M}. However, the utilization of the dual-path method introduces a significant processing overhead due to the presence of overlapped chunks and inefficient modelling of global information through inter-processing \cite{Hyunseok2023L}. In our prior study \cite{Zhao2023M}, we introduced MossFormer, an alternative approach involving joint attention. This method employs non-overlapped chunks and linearized attention across the entire sequence to capture global information, leading to notable performance improvements. Nonetheless, the Transformer-based strategy places greater emphasis on long-range coarse-scale dependencies, while paying less attention to the finer-scale recurrent patterns. As a result, the Transformer-only solution remains sub-optimal. More recently, numerous efforts have been undertaken to enhance the Transformer-based approach, including the quasi-dual-path network (QDPN) \cite{rixen22_interspeech}, the SFSRNet \cite{Rixen2022M}, and the Separate and Diffuse \cite{lutati2023separate}. However, the advancements they introduce invariably lead to either considerably larger models or more intricate composite models.  

\vspace{-1.0mm}
In this work, we enhance the Transformer-based approach through the introduction of an innovative RNN-free recurrent module. 
This module is seamlessly integrated into our MossFormer model, as depicted in Fig. \ref{fig1}, for 
complementing the MossFormer's capabilities in modelling finer-scale recurrent patterns, a crucial aspect in the speech separation task.  Our recurrent module is built upon the feedforward sequential memory network (FSMN) \cite{zhang2018deepfsmn}, which leverages a feedforward structure with a memory component. We enhance the FSMN by incorporating dilations to achieve broader receptive fields. Moreover, we implement gated convolutional units (GCU) and dense connections to further expand the dilated FSMN's capabilities while permitting dimensionality reduction. By combining memory blocks, gated units, and densely dilated convolutions, we provide an effective mechanism for capturing finer-scale recurrent patterns in  sequential information without requiring explicit recurrent connections in RNNs. 
This upgraded hybrid model is dubbed MossFormer2. Our experimental results demonstrated that substantial performance enhancements are achieved. MossFormer2 has not only set new state-of-the-art benchmarks on WSJ0-2mix/3mix, Libri2Mix but also demonstrated remarkable results on the more challenging WHAM!/WHAMR! benchmarks. 
 
\section{The MossFormer2 Model}
\begin{figure}[t]
  \centering
  \includegraphics[width=8.5cm]{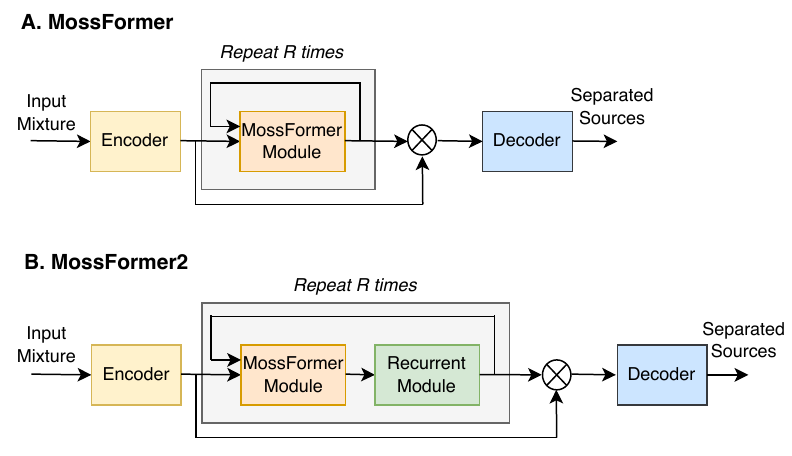}
  \vspace{-3mm}
  \caption{(A) The diagram of the base MossFormer model. (B) The diagram of the proposed MossFormer2 model. Compared to MossFormer using only MossFormer module, hybrid MossFormer and recurrent modules are adopted in MossFormer2.}
  \label{fig1}
\vspace{-3mm}
\end{figure}

Given the speech mixture $x=\sum_{i=1}^{C}s_i$, our objective is to use a deep learning model to estimate $C$ individual sources $s_i\in\mathbb{R}^{1xT}, i=1,2,\dots,C$. Our proposed model, MossFormer2, adopts the time-domain masking-net framework \cite{Luo2019N} which still retains the encoder-separator-decoder framework utilized in MossFormer. However, it focuses on enhancing the separator component, as illustrated in Fig. \ref{fig1}. The encoder-decoder structure is responsible for feature extraction and waveform reconstruction. The masking-net then maps the encoded output to a set of masks.
 
\subsection{Encoder and Decoder}
Following the approach of MossFormer, the encoder utilizes a one-dimensional (1-D) convolutional layer (Conv1D) combined with a rectified linear unit (ReLU) to encode the input mixture waveform $x\in\mathbb{R}^{1xT}$. This encoding process transforms the waveform into a non-negative speech embedding sequence $X\in\mathbb{R}^{N\times S}$, with $N$ representing the embedding dimension and $S$ denoting the length of the encoded sequence. Assuming a kernel size of $K_1$ and a stride of $K_1/2$ for the encoder, the encoded sequence length is calculated as $S=2T/K_1-1$. On the other hand, the decoder reconstructs the original waveform of length $T$ using a transposed 1-D convolutional layer with the same kernel size and stride as the encoder.


\subsection{Hybrid MossFormer and Recurrent Modules}
\begin{figure}[t]
  \centering
  \includegraphics[width=8.0cm]{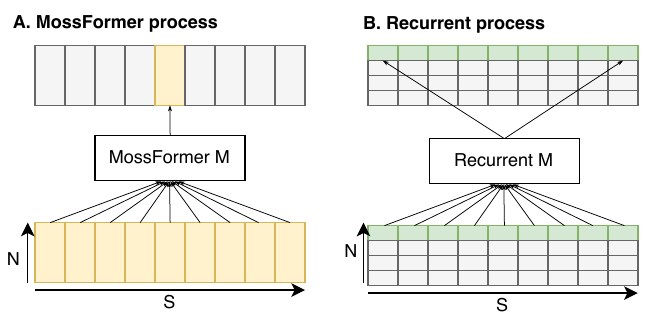}
  \vspace{-3mm}
  \caption{Illustration of hybrid MossFormer and recurrent modules. In this setup, the MossFormer module (A) handles the complete sequence to capture global dependencies, whereas the recurrent module (B) engages in recurrent learning across each embedding dimension. ($S$: the  sequence length, $N$: the embedding dimension).}
  \label{fig2}
\vspace{-3mm}
\end{figure}

The MossFormer module, as illustrated in Fig. \ref{fig1}, remains consistent across both MossFormer and MossFormer2. The core concept of the MossFormer module involves applying self-attention \cite{Vaswani2017N} to the entire sequence. For efficient self-attention across the extensive sequence, we adopt the joint local-global self-attention strategy in MossFormer. This approach concurrently executes full-computation self-attention on non-overlapping local segments and employs a linearized, resource-efficient self-attention mechanism over the entire sequence. Illustrated in Fig. \ref{fig2} (Left), this self-attention mechanism enables the MossFormer module to learn each subsequent layer's embedding by considering all elements within the current layer's sequence. 
Not relying on recurrence, the self-attention primarily captures long-range, coarse-scale dependencies.
However, speech signals inherently exhibit recurrent patterns that manifest in phonetic structures, prosody, and semantic associations, all of which play a significant role in speech separation.
Consequently, relying solely on self-attention mechanisms is inadequate for the formulation of an effective speech separation model.

In MossFormer2, we introduce a dedicated recurrent module to model intricate temporal dependencies within speech signals. We hypothesize that distinct embedding levels retain distinct recurrent patterns, thus this recurrent module conducts recurrent learning on each embedding dimension, as depicted in Fig. \ref{fig2} (Right). All the embedding dimensions can undergo parallel learning. Leveraging the combined strengths of self-attention and recurrent modeling, MossFormer2 facilitates the capture of both broad dependencies and localized recurrent patterns.

\subsection{RNN-Free Recurrent Module}
\begin{figure*}[t]
  \centering
  \includegraphics[width=17cm]{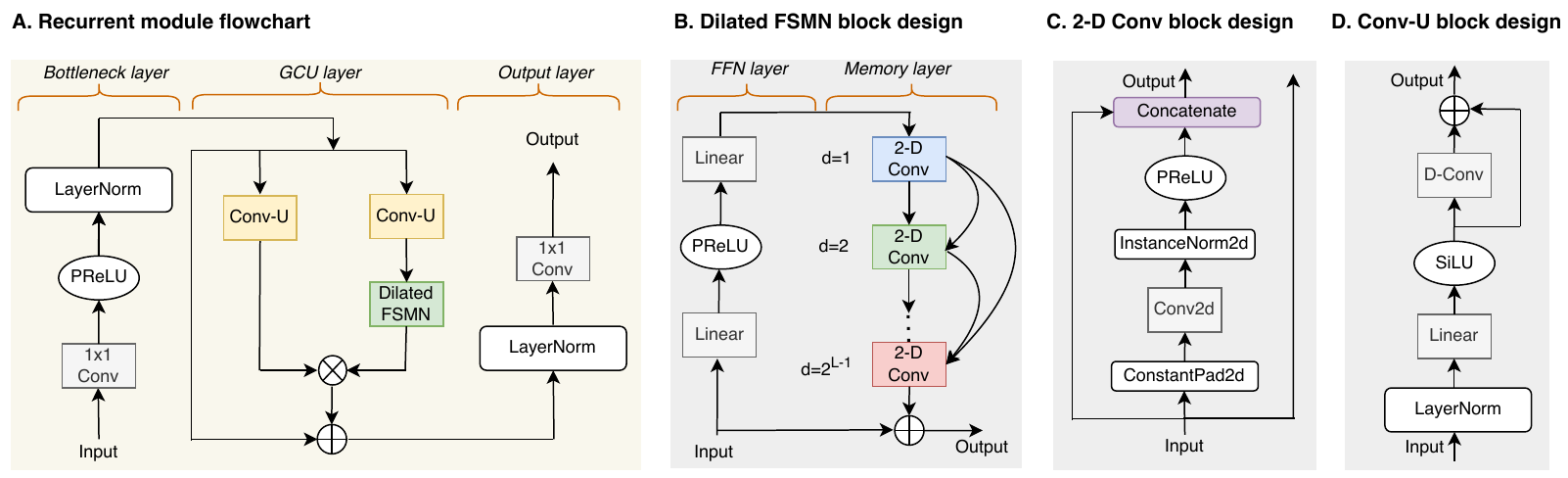}
  \vspace{-3mm}
  \caption{(A) The flowchart of the proposed recurrent module. It comprises a bottleneck layer, a GCU layer, and an output layer. The GCU layer is built on gated convolutional units and a dillated FSMN block. (B) The design of the dilated FSMN block. (C) The design of the 2-D Conv block. (D) The design of the Conv-U block.}
  \label{fig3}
\vspace{-3mm}
\end{figure*}

While LSTM and GRU architectures are widely used for recurrent tasks, they come with notable drawbacks due to sequential processing with recurrent connections. We introduce a recurrent module based on FSMN without recurrent connections, illustrated in Fig. \ref{fig3}.

The proposed recurrent module is composed of a bottleneck layer, a GCU layer, and an output layer. Within the GCU layer, we incorporate a dilated FSMN block (Dilated FSMN) along with two convolutional units (Conv-U). This dilation extension enables the FSMN to cover broader receptive fields while requiring fewer memory resources. The GCU architecture is employed, inspiring from the effectiveness of the gating mechanism in GLU \cite{dauphin2017language}. In order to further improve the dilated FSMN and allow for a decreased embedding dimensionality, we propose to use convolutional units in place of linear units. This is accomplished by initially implementing a bottleneck layer through a $1\times1$ convolutional layer, which is followed by a PReLU activation \cite{He2015Z} and a LayerNorm layer. The purpose of the bottleneck layer is to decrease the embedding dimensionality while retaining crucial features. The resulting output of the bottleneck layer is then passed through the GCU layer. 

Concretely, the GCU layer maps a sequence $X\in\mathbb{R}^{N^\prime\times S}$ to an output $O\in\mathbb{R}^{N^\prime\times S}$ as
\begin{align}
&U=\mathrm{Conv\_U }(X) \in\mathbb{R}^{N^\prime\times S},\quad V=\mathrm{Conv\_U }(X) \in\mathbb{R}^{N^\prime\times S}\\
&Y=\mathrm{Dilated\_FSMN }(V) \in\mathbb{R}^{N^\prime\times S} \\
&O=X + (U\otimes Y)
\end{align}
where $\otimes$  denotes element-wise multiplication and $N^\prime$ the reduced embedding dimensionality after the bottleneck layer. To facilitate model training, we add a skip connection to link the GCU layer's input to its output. The output layer employing a LayerNorm followed by a $1\times1$ convolution layer, restores the embedding dimensionality. 
 
\subsubsection{Dilated FSMN Block}
The dilated FSMN block comprises a feed-forward (FFN) layer and a memory layer. While the FFN layer maintains the consistent structure of the original FSMN, we enhance the memory layer by employing stacked two-dimensional (2-D) dilated convolutional (2-D Conv) blocks, as depicted in Fig. \ref{fig3} (B). The dilation factors $1, 2, \ldots, 2^{L-1}$ are employed to increase the receptive field and facilitate context aggregation at varying resolutions. The 2-D convolution is chosen for its ease of implementing the grouping mechanism. 
 The speech embedding from the FFN layer, which is initially in 2-D, is reshaped into a three-dimensional (3-D) sequence. This converts the embedding dimension into the input channels. These input channels are subsequently divided into distinct groups, with each group undergoing convolution using a dedicated set of filters. 
 
Within the memory layer, we extend the utilization of feed-forward dense connections to establish interconnections among each 2-D Conv block and all other 2-D Conv blocks, as introduced in \cite{huang2018densely}. This extension aims to enhance the information flow and facilitate the gradient propagation. To elaborate further, the $l^{th}$ block is supplied with speech embeddings from all preceding blocks, namely $X_0, X_1, \ldots, X_{l-1}$, as its input:
\begin{equation}
X_l = H_l\left(\left[X_0,X_1,...,X_{l-1}\right]\right)
\end{equation}  
where $\left[X_0,X_1,...,X_{l-1}\right]$ refers to the concatenation of the inputs in blocks $0,...,l-1$, and $H_l$ denotes the composite function of operations in 2-D Conv as illustrated in Fig. \ref{fig3} (C). The 2-D Conv block comprises constant padding, a 2-D convolutional layer, an instance normalization \cite{ulyanov2017instance}, and a PReLU activation. The output is concatenated with the input to facilitate the dense connection. Therefore, the dilated FSMN employs only feed-forward connections along with the convolutional memory layer to capture recurrent patterns.


\subsubsection{Conv-U Block}
The structure of the Conv-U block as shown in Fig. \ref{fig3} (D) follows the convolution module presented in the MossFormer work. Within this block, the sequence initially undergoes a LayerNorm layer, followed by a linear layer and a SiLU activation \cite{elfwing2017sigmoidweighted}. Subsequently, a 1-D depthwise convolution (D-Conv) layer is applied. The skip-connections are used to facilitate model training. The role of the Conv-U block is to assist the GCU layer within the recurrent module in capturing position-wise local patterns.

\section{Experiments}
\subsection{Dataset}
We evaluated the performance of MossFormer2 on various speech separation benchmarks, including WSJ0-2/3mix \cite{Hershey2016Z}, Libri2Mix \cite{cosentino2020librimix}, and WHAM!/WHAMR! \cite{Wichern2019J, Maciejewski2019G} datasets. The WSJ0-2/3mix dataset comprises 2 and 3 speaker clean mixtures, with speeches randomly drawn from the LDC WSJ-0 corpus. This dataset contains 30 hours of training, 10 hours of validation, and 5 hours of evaluation data. WHAM! represents a noisy variant of WSJ0-2mix, while WHAMR! is a reverberant version of WHAM!. Additionally, Libri2Mix involves clean mixtures of 2 speakers, utilizing speeches from the LibriSpeech ASR corpus \cite{Panayotov2015}. This dataset comprises 106 hours of training, along with 5.5 hours each of validation and evaluation data. The 8 kHz versions of the datasets are employed, with dynamic mixing  data augmentation \cite{Subakan2021M} applied to all datasets except the larger Libri2Mix dataset.

\subsection{Training Setup}
We built our models using the SpeechBrain toolkit \footnote{https://github.com/speechbrain/speechbrain} and optimized them based on the SI-SDR training loss \cite{Roux2019S}. Our training takes place on a single NVIDIA V100 GPU with 32 GB of memory. Each model undergoes up to 200 epochs of training, utilizing the Adam optimizer \cite{Kingma2014J} with an initial learning rate set at $15e^{-5}$ and a batch size of 1.
During the training process, the learning rate remains constant for 85 epochs, after which it is reduced by a factor of 0.5. For gradient control, we applied clipping, capping $l_2$ norm of training gradients at 5. We have made audio samples available online\footnote{https://github.com/alibabasglab/MossFormer2}, and plan to release the source code at a later time.
\begin{table}
\center
\footnotesize
\caption{Comparison for MossFormer and MossFormer2 on the WSJ0-2mix dataset. RTF denotes the real-time factor on test set.}
\setlength\tabcolsep{5.0pt} 
\begin{tabular}{lcccccccc}
\specialrule{.1em}{.05em}{.05em}
Model                 &Para.(M) & $R$ & $N$ & $K$ &$N^\prime$ & $L$ &SI-SDRi&RTF\\ \hline 
MossFormer (S)       &25.3   & 25  & 384 &16 &- &- &22.5 & 0.025    \\ 
MossFormer           &42.1  & 24  & 512 &16 &- &- &22.8  & 0.038       \\ 
MossFormer2 (S)      &37.8   & 25  & 384 &16 &256 &2 &23.2 & 0.036    \\ 
MossFormer2          &55.7   & 24  & 512 &16 &256 &2 &\textbf{24.1} &0.053   \\ \hline
\specialrule{.1em}{.05em}{.05em}
\label{tab1}
\end{tabular}
\end{table}

\begin{table}
\vspace{-6mm}
\center
\footnotesize
\caption{Performance comparison of MossFormer2 with the other state-of-the-art speech separation models on the WSJ0-2/3mix and Libri2Mix benchmark datasets.}
\setlength\tabcolsep{5.0pt} 
\begin{tabular}{lcccc}
\specialrule{.1em}{.05em}{.05em}
\multirow{2}{*}{Model}  &\multirow{2}{*}{Para.(M)} & \multicolumn{3}{c}{SI-SDRi}         \\
\cline{3-5}
                                         &        & WSJ0-2mix &WSJ0-3mix  & Libri2Mix        \\ \hline 
Conv-TasNet \cite{Luo2019N}          & 5.1    & 15.3  & 12.7    & 14.7               \\ 
DPRNN \cite{Luo2020Z}                & 2.6    & 18.8  & 14.7    & \quad -\quad\          \\
VSUNOS \cite{Nachmani2020Y}          & 7.5    & 20.1  & 16.9    &  \quad -\quad\         \\ 
DPTNet \cite{Chen2020Q}              & 2.6    & 20.2  & \quad -\quad\ & \quad -\quad\  \\ 
Wavesplit  \cite{Zeghidour2021D}     & 29     & 22.2  &17.8     & 19.5 \\ 
SepFormer \cite{Subakan2021M}        & 25.7   & 22.3  &19.5     & 19.2 \\ 
QDPN \cite{rixen22_interspeech}      & 200.0  & 23.6  &\quad -\quad\ &\quad -\quad\ \\
Separate And Diffuse \cite{lutati2023separate}   & \quad -\quad\ &23.9 &20.9 &\quad -\quad\ \\
SFSRNet \cite{Rixen2022M}                           & 59.0   & 24.0  &\quad -\quad\ &20.4 \\ \hline
MossFormer                           & 42.1   & 22.8  & 21.2    &  19.7\ \\  
MossFormer2                          & 55.7   & \textbf{24.1}  & \textbf{22.2}    & \textbf{21.7}       \\ \hline 
\specialrule{.1em}{.05em}{.05em}
\label{tab2}
\end{tabular}
\vspace{-6mm}
\end{table}

\subsection{Results}
We used SI-SDR improvement (SI-SDRi) as the evaluation metric. The effectiveness of incorporating the recurrent module into MossFormer2, as opposed to the base MossFormer model, is highlighted in Table \ref{tab1}. With the addition of the recurrent module, 
MossFormer2 results in a notable rise in the SI-SDRi score from 22.8 dB to 24.1 dB on the WSJ0-2mix dataset.
To verify that the performance boost is not solely attributed to the increase in parameters, we created scaled-down versions of MossFormer (S) and MossFormer2 (S). Notably, while expanding the model size from MossFormer (S) to MossFormer led to only marginal performance improvement, MossFormer2 (S) outperforms MossFormer even with fewer parameters, achieving a 0.4 dB absolute improvement in SI-SDRi. This experiment emphasizes that the recurrent module complements the MossFormer module in enhancing speech separation capabilities.  

To further demonstrate the time efficiency of the proposed recurrent module without using RNN, we calculated the real-time factor (RTF) for both MossFormer and MossFormer2, along with their scaled-down versions, using the WSJ0-2mix test set. These measurements were conducted on the same single GPU card used for training. The results reveal that MossFormer2 (S) not only achieves higher SI-SNRi scores compared to MossFormer but also exhibits a smaller RTF. While MossFormer2 has a slight increase in RTF compared to MossFormer, the RTF range remains at a notably low level.

Table \ref{tab2} presents the outcomes on the WSJ0-2mix/3mix and Libri2Mix datasets. MossFormer2 was evaluated against the most promising benchmarks reported in the literature. We observed that the  quasi-dual-path approach has advanced QDPN over SepFormer. Nonetheless, QDPN substantially increases model size to 200 million parameters. In contrast, MossFormer2 surpasses QDPN with nearly a quarter of the parameters. While SFSRNet provides competitive SI-SDRi scores, it concentrates on Super-Resolution network aspects within the frequency reconstruction context, necessitating a complex multi-loss training strategy. In contrast, our MossFormer2 design exploits on the recurrent nature of speech signals and focuses on the recurrent module, employing a much simpler training strategy. Notably, MossFormer2 outperforms Separate And Diffuse, even though it incorporates a pre-trained vocoder, DiffWave, in addition to SepFormer. Furthermore, MossFormer2 achieves superior SI-SDRi scores on Libri2Mix in the absence of dynamic mixing.
In Table \ref{tab3}, the results for the WHAM! and WHAMR! datasets are presented. It is observed that QDPN yields minor enhancements compared to SepFormer on WHAMR!. On the other hand, MossFormer2 not only exhibits significant improvements over MossFormer but also outperforms QDPN and other methods by a large margin.  

\begin{table}
\center
\footnotesize
\caption{Performance comparison of MossFormer2 with the other state-of-the-art speech separation models on the WHAM! and WHAMR! benchmark datasets.}
\begin{tabular}{lccc}
\specialrule{.1em}{.05em}{.05em}
\multirow{2}{*}{Model}  &\multirow{2}{*}{Para.(M)} & \multicolumn{2}{c}{SI-SDRi}         \\
\cline{3-4}
                                     &        & WHAM! &WHAMR!        \\ \hline 
Conv-TasNet \cite{Luo2019N}          & 5.1    & 12.7  & 8.3                   \\ 
DPRNN \cite{Luo2020Z}                & 2.6    & 13.9  & 10.3              \\
VSUNOS \cite{Nachmani2020Y}          & 7.5    & 15.2  & 12.2            \\ 
Wavesplit  \cite{Zeghidour2021D}     & 29     & 16.0  & 13.2     \\ 
SepFormer \cite{Subakan2021M}        & 25.7   & 16.4  & 14.0     \\ 
QDPN \cite{rixen22_interspeech}                              & 200.0  & \quad -\quad\  & 14.4  \\ \hline
MossFormer                           & 42.1   & 17.3  & 16.3    \\  
MossFormer2                          & 55.7   & \textbf{18.1}  & \textbf{17.0}           \\ \hline 
\specialrule{.1em}{.05em}{.05em}
\label{tab3}
\end{tabular}
\end{table}

\subsection{Ablation Studies}
Ablation studies were conducted on MossFormer2 to analyze the impact of each introduced element, as detailed in Table \ref{tab4}. The findings indicate that both dilations and dense connections contribute to enhancing FSMN's capabilities. Furthermore, the Conv\_U block, along with the gated mechanism, benefits the GCU layer more effectively compared to a linear layer. It was observed that the absence of the bottleneck layer and output layer leads to a decline in performance, demonstrating the effectiveness of   information control by applying the bottleneck layer. 

\begin{table}
\vspace{-6mm}
\center
\footnotesize
\caption{Ablation studies for MossFormer2 on the dilated FSMN, the GCU layer, and the bottleneck and output layers.}
\begin{tabular}{lc}
\specialrule{.1em}{.05em}{.05em}

Model                                                          & SI-SDRi        \\ \hline 
MossFormer2                                                      &\textbf{24.1}             \\ 
Without dilation in FSMN                                         &23.9              \\ 
Without dense connections in Dilated FSMN                        &24.0                  \\ 
Replace Conv\_U with Linear in the GCU layer                      &23.8          \\
Remove convolutional units (Conv\_U) from the GCU layer           &23.5    \\
Remove bottleneck and output layers  from the recurrent module   &23.9    \\ \hline 
\specialrule{.1em}{.05em}{.05em}
\label{tab4}
\end{tabular}
\vspace{-6mm}
\end{table}

\section{Conclusions}
We introduced MossFormer2, a hybrid Transformer and recurrent model for monaural speech separation. Unlike prior approaches, our model captures both recurrent patterns and global dependencies within a single framework. By incorporating dilations, dense connections, and a gated mechanism, our RNN-free recurrent module significantly improves separation performance. Our study highlights the impact of each technique and showcases MossFormer2's superior performance over MossFormer and other state-of-the-art models on diverse benchmarks.
\bibliographystyle{IEEEbib}
\bibliography{refs}

\end{document}